\documentclass[twocolumn]{aastex631}

\usepackage{amsmath}
\usepackage{CJK}

\newcommand{\fargo}{ FARGO3D}

\newcommand{\pmass}{q}
\newcommand{\ar}{h_\mathrm{0}}
\newcommand{\deltavr}{\delta v_{\theta}(r)}
\newcommand{\deltav}{\Delta\delta v_{\theta}}

\newcommand{\tanhf}{Tanh}
\newcommand{\stanf}{Stan}

\definecolor{turquoise}{cmyk}{0.65,0,0.1,0.3}
\definecolor{purple}{rgb}{0.65,0,0.65}
\definecolor{dark_green}{rgb}{0, 0.5, 0}
\definecolor{orange}{rgb}{0.9, 0.6, 0.1}
\definecolor{red}{rgb}{0.8, 0.2, 0.2}
\definecolor{darkred}{rgb}{0.6, 0.1, 0.05}
\definecolor{blueish}{rgb}{0.0, 0.3, .6}
\definecolor{light_gray}{rgb}{0.7, 0.7, .7}
\definecolor{pink}{rgb}{1, 0, 1}
\definecolor{greyblue}{rgb}{0.25, 0.25, 1}

\graphicspath{{./}{attachment/}}

\begin{document}
\begin{CJK*}{UTF8}{gbsn}
\title{PPDONet: Deep Operator Networks for Fast Prediction of Steady-State Solutions in Disk-Planet Systems}

\author[0009-0006-0889-3132]{Shunyuan Mao (毛顺元)}
\affiliation{Department of Physics and Astronomy, University of Victoria, Victoria, BC V8P 5C2, Canada, symao@uvic.ca, rbdong@uvic.ca}

\author[0000-0001-9290-7846]{Ruobing Dong (董若冰)}
\affiliation{Department of Physics and Astronomy, University of Victoria, Victoria, BC V8P 5C2, Canada, symao@uvic.ca, rbdong@uvic.ca}

\author[0000-0002-5476-5768]{Lu Lu}
\affiliation{Department of Chemical and Biomolecular Engineering, University of Pennsylvania, Philadelphia, PA 19104, USA}

\author[0000-0001-9036-3822]{Kwang Moo Yi}
\affiliation{Department of Computer Science, University of British Columbia, Vancouver, BC V6T 1Z4, Canada}

\author{Sifan Wang}
\affiliation{Graduate Group in Applied Mathematics and Computational Science, University of Pennsylvania, Philadelphia, PA 19104, USA}

\author[0000-0002-2816-3229]{Paris Perdikaris}
\affiliation{Department of Mechanical Engineering and Applied Mechanics, University of Pennsylvania, Philadelphia, PA 19104, USA}

\begin{abstract}
We develop a tool, which we name Protoplanetary Disk Operator Network (PPDONet), that can predict the solution of disk-planet interactions in protoplanetary disks in real-time.
We base our tool on Deep Operator Networks (DeepONets), a class of neural networks capable of learning non-linear operators to represent deterministic and stochastic differential equations.
With PPDONet we map three scalar parameters in a disk-planet system -- the Shakura \& Sunyaev viscosity $\alpha$, the disk aspect ratio $\ar{}$, and the planet-star mass ratio $\pmass{}$ -- to steady-state solutions of the disk surface density, radial velocity, and azimuthal velocity.
We demonstrate the accuracy of the PPDONet solutions using a comprehensive set of tests.
Our tool is able to predict the outcome of disk-planet interaction for one system in less than a second on a laptop.
A public implementation of PPDONet is available at \url{https://github.com/smao-astro/PPDONet}.

\end{abstract}


\section{Introduction}\label{sec: intro}
\end{CJK*}
Planets form in the protoplanetary disks surrounding newborn stars. As they do, planets interact via gravity with the host disk and produce large-scale structures, such as gaps, spiral arms, and dust clumps. 
This interaction, which can be represented via fluid dynamics, is commonly referred to as disk-planet interaction \citep{kley2012planet}. 
Simulations of protoplanetary disk evolution and disk-planet interaction are essential, for example, to understand how disks accrete and disperse \citep{tabone2022mhd}, to interpret observed disk structures \citep{dong2015observational2,dong2017mass,liu2018new}, to constrain planet properties \citep{fung2014empty,fung2015inferring,zhang2018disk}, and to study the orbital evolution of planets \citep{paardekooper2022planet}. 

Numerical methods have been developed for decades to simulate disk evolution and disk-planet interaction \citep{paardekooper2022planet}.
However, such methods typically require a large amount of computing.
For example, simulating a 2D disk with $270$ ($r$) by $810$ ($\theta$) resolution for $2,000$ orbits takes $20$ GPU hours \citep{fung2014empty}. 
As a result, modeling a large number of disk-planet systems, or exploring the parameter space of their interaction can be expensive. 
Moreover, as the number of observed disks grows fast \citep{benisty2022optical,bae2022molecules}, modeling all of these observations using simulations is becoming impractical.

There are multiple reasons why massive disk simulations are so expensive to perform. Among them two are critical.
First, each numerical simulation is run from scratch. 
In other words, even though much of the simulation outcome may be similar for a small change in the simulation configuration, there is no reuse of these similar simulation outcomes. Thus, the computing cost scales linearly with the number of simulations executed.
Second, each simulation itself is expensive. 
Disk evolution and disk-planet interaction simulations often need to be run for many numerical time steps, with each length dictated by the Courant--Friedrichs--Lewy condition to guarantee accuracy.
A typical disk-planet system may evolve thousands of orbits before reaching a steady state, equivalent to millions of time steps. 
In this work, we tackle both of these issues with the help of Machine Learning (ML), which then allows us to ``learn'' to utilize similar simulations, and at the same time eliminate the need for time-stepping execution.

Recently, ML techniques have been applied to model disk-planet systems, although not for the same problem as we are interested in---to predict the surface density and velocity fields of a disk-planet system. 
\cite{auddy2020machine} used a deep neural network to predict planet mass based on 1D surface density profiles of gaps.
In a follow-up work, \cite{auddy2021dpnnet} developed a convolutional neural network (CNN) to decipher planet mass from 2D dust density distribution.
\cite{zhang2022pgnets} developed a similar neural network to infer planet mass from synthetic observations.
However, these methods are focused on constraining scalar outputs from disk morphologies, rather than predicting physical quantity distributions.

Thus, when applying ML to simulating disk evolution, care must be taken in how one chooses the neural network architecture.
We opt for a recent architecture, DeepONets~\citep{lu2021learning}, that greatly reduces the computing cost of simulating physical systems.
In particular, they learn non-linear operators that represent deterministic and stochastic differential equations.
They can predict solutions for differential equations with parametric boundary conditions, initial conditions, or forcing terms.
As a result, they provide better generalization and faster convergence with respect to the volume of training data~\citep{lu2021learning}.

In this inaugural work in our series of applying machine learning to model protoplanetary disks, we introduce a new tool based on DeepONets that can instantaneously predict the steady-state disk structure in both surface density and velocities of a disk-planet system.
We train DeepONets (\S \ref{sec: implementation}) with hydrodynamic simulations produced using numerical solvers and test the networks by reproducing empirical relationships in disk-planet interactions (\S \ref{sec: tests}).
This tool, PPDONet, is publicly available on GitHub\footnote{PPDONet codebase: \url{https://github.com/smao-astro/PPDONet}} under a GPL v3.0 License and version 0.1.0 is archived in Zenodo \citep{zenodo}. 
A web interface\footnote{Web app: \url{https://ppdonet-1.herokuapp.com}} has been developed for the convenience of usage as well.

\section{DeepONets}\label{sec: deeponet}
\begin{figure*}[htb!]
    \centering
    \includegraphics[trim=0 0 0 0, clip,width=1\textwidth,angle=0]{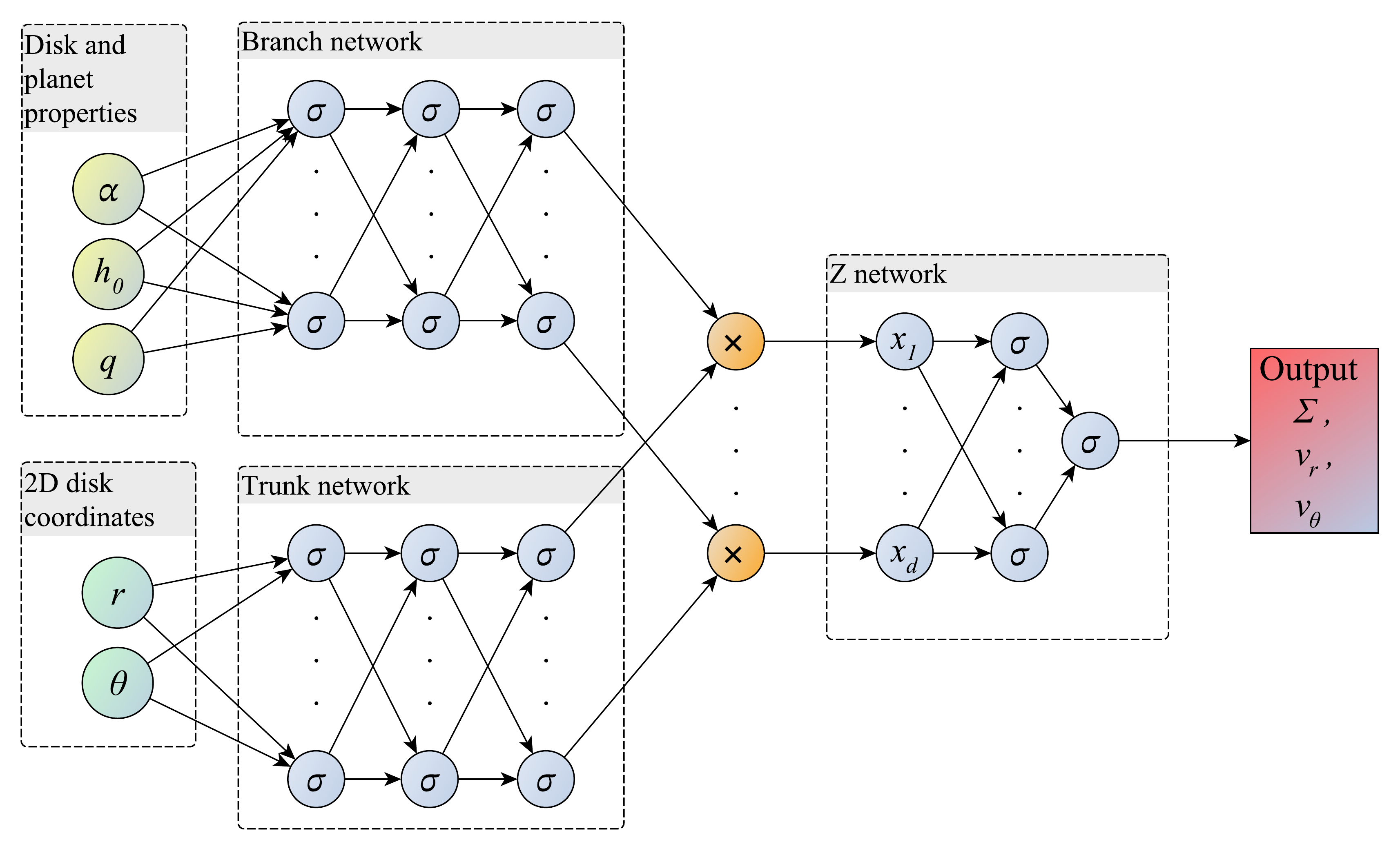}
    \caption{PPDONet architecture used in this work. A fully connected subnetwork, called ``branch network'', encodes scalar parameters, the disk and planet properties, while another fully connected subnetwork, called ``trunk network'', encodes the coordinates. We produce the inputs $x_i$, $i=1,...,d,\ d=50$, for the third network (``Z network'') by element-wise multiplication of the outputs from the branch and trunk networks. The Z network is a fully connected single-hidden-layer network whose output represents surface density or velocity. The $\sigma$ in blue nodes are neurons in layers.}
    \label{fig: zdeeponet arch}
\end{figure*}
As our work utilizes the DeepONet~\citep{lu2021learning} architecture, we first review DeepONet.
DeepONet is a machine learning-based approach that utilizes a neural network to approximate a mathematical operator denoted as $G$. In disk-planet systems, $G$ maps a set of scalar parameters, $\vec{p}$, to a distribution function, $G(\vec{p})$, that can be evaluated at any desired location. Specifically, in our networks $\vec{p}$ includes the Shakura \& Sunyaev viscosity $\alpha$, the disk aspect ratio $\ar{}$, and the planet-star mass ratio $\pmass{}$, and $G(\vec{p})\left(r,\theta\right)$ represents the surface density or velocity at a specified point $(r,\theta)$. Therefore, the DeepONet architecture comprises two kinds of inputs: scalar parameters, $\vec{p}$, and coordinates $(r,\theta)$, and one output, which can be either surface density, radial velocity, or azimuthal velocity.

Figure \ref{fig: zdeeponet arch} depicts the architecture of PPDONet. To map inputs ($\vec{p}$) to solutions ($G(\vec{p})$), the scalar parameters are encoded by a subnetwork referred to as the ``branch network.'' This network is fully connected and consists of four $100$-neuron hidden layers ($4\times100$). Concurrently, the coordinate inputs are encoded by another fully connected subnetwork, the ``trunk network,'' which has five layers of $256$ neurons each. We obtain the inputs $x_i$, $i=1,...,d,\ d=50$, for the third network (``Z network'' in Figure \ref{fig: zdeeponet arch}) by element-wise multiplication of the outputs from the branch and trunk networks. The Z network is a fully connected single-hidden-layer network whose output represents surface density or velocity.
We employ the self-scalable \tanhf{} activation function (\stanf{}) \citep{gnanasambandam2022self} for fitting surface density and azimuthal velocity, as it enables fast convergence and results in small training and testing errors; however, the \tanhf{} activation function outperforms \stanf{} for learning radial velocity.

\section{Implementation}\label{sec: implementation}
    \subsection{Hydrodynamic simulations for training, validating, and testing neural networks}\label{sec: hydro simulations}
We focus on large-scale structures, such as gaps and spiral arms, induced by a single planet on a fixed circular orbit in gaseous disks. 
The disk's initial profiles are described by the equations:
\begin{subequations}
    \begin{eqnarray}
        \Sigma&=&\Sigma_\mathrm{0}\left(r / r_{\mathrm{p}}\right)^{-1 / 2},\\
        v_r&=&-\frac{3}{2} \alpha{\ar{}}^2 \sqrt{\frac{G M_\odot\left(1+\pmass{}\right)}{r}},\\
        v_{\theta}&=&\sqrt{1-\frac{3}{2}{\ar{}}^2} \sqrt{\frac{G M_\odot\left(1+\pmass{}\right)}{r}}\label{eq: v_theta_ic},
    \end{eqnarray}
\end{subequations}
where $\Sigma$ is the surface density, $v_r$ is the radial velocity, $v_{\theta}$ is the azimuthal velocity, $\alpha$ is the Shakura \& Sunyaev  viscosity, and $\ar{}$ is the disk aspect ratio. Both $\alpha$ and $\ar{}$ are kept constant throughout the disk. 
The models are developed for a fixed initial radial surface density profile and a radially constant disk aspect ratio.
The planet is at $r=r_\mathrm{p}$ with mass $m_{\mathrm p}=\pmass{}M_\odot$. The initial surface density at $r_\mathrm{p}$ is $\Sigma_\mathrm{0}=1$.  
This setup ensures our model corresponds to planet-free steady-state accretion disks \citep{fung2014empty}.
The boundary conditions are fixed and determined by initial values, which ensure a constant mass inflow \citep{fung2014empty}.

\begin{deluxetable}{ccc}
\tablecaption{Disk and planet parameter space\label{table: parameter space}}
\tablehead{\colhead{Parameter} & \colhead{Minimum} & \colhead{Maximum}
} 
\startdata
$\alpha$ & $3\times10^{-4}$ & $0.1$ \\
$h_\mathrm{0}$ & $0.05$ & $0.1$ \\
$\pmass{}$ & $5\times10^{-5}$ & $2\times10^{-3}$ \\
\enddata
\end{deluxetable} %

The outcomes of disk-planet interaction in our simulations are determined by three parameters: $\alpha$, $\ar{}$, and $\pmass{}$. Massive planets open deep gaps, while large viscosity and aspect ratio hinder the opening of deep gaps. With $\pmass{}>2\times10^{-3}$, or $\alpha<3\times10^{-4}$, a disk may develop vortices or other asymmetric and time-varying structures \citep{fung2014empty}. In this work, we only focus on disks capable of reaching a steady state with parameters bounded by Table \ref{table: parameter space}.

To collect data for training neural networks, choose the best ML models, and test their performance on unseen parameters, we generate $768$ \fargo{} \citep{masset2000fargo,benitez2016fargo3d} hydrodynamic simulations with $r\times\theta=381\times1143$ resolution.
We sample $\alpha$, $\ar{}$, and $\pmass{}$ from Sobol sequences \citep{sobol1967distribution}, quasi-random low-discrepancy sequences effective in generating inputs for machine learning tasks \citep{wu2023comprehensive}.
We divide the $768$ FARGO3D simulations into three groups with $448$, $64$, and $256$ cases for training, validation, and testing, respectively. In training, we compare neural network predictions with simulations to formalize loss functions. In testing, we measure errors on unseen simulations to assess the generalization of our neural networks.
All simulations are run for $0.314\tau_{\mathrm{\nu}}$ to reach (quasi) steady state, where $\tau_{\mathrm{\nu}}$ is the disk viscous timescale
\begin{equation}
    \tau_{\mathrm{\nu}}\approx\frac{r_\mathrm{p}^2}{\nu}=\frac{1}{\alpha c_{\mathrm{s}} h}=\frac{1}{\alpha {h_0}^2}.
\end{equation} 

     \subsection{Network training}\label{sec: network training}
To fit the steady-state solution of three quantities -- surface density $\Sigma$, radial velocity $v_r$, and azimuthal velocity $v_{\theta}$ -- we train three separate neural networks. As some of the inputs span several orders of magnitude, we convert them to logarithmic scales. We then normalize both the scalar parameters and coordinate inputs. To facilitate learning, we perform two additional steps: 1) we take the logarithm of the surface density, and 2) we subtract the background from the radial and azimuthal velocities.
As a result, our loss function for surface density is
\begin{equation}
    L_\mathrm{\Sigma} = \frac{1}{N} \sum\limits_{i=1}^N \left[\log(\Sigma_i^{\mathrm{pred}}) - \log(\Sigma_i^{\mathrm{truth}})\right]^2 \label{eq: sigma loss},
\end{equation}
while our loss functions for velocities are
\begin{equation}
    L_{v_r} = \frac{1}{N} \sum\limits_{i=1}^N \left[{v_r}_i^{\mathrm{pred}} - {v_r}_i^{\mathrm{truth}}\right]^2
\end{equation}
and
\begin{equation}
    L_{v_\theta} = \frac{1}{N} \sum\limits_{i=1}^N \left[{v_\theta}_i^{\mathrm{pred}} - {v_\theta}_i^{\mathrm{truth}}\right]^2.
\end{equation}
The indices $i$ go through all the grid points in all the simulations in a training batch \footnote{To compare with \fargo{}, the grid points for surface density are cell-centered, while velocities are face-centered \citep[see][\S2.3]{benitez2016fargo3d}}.
We regard FARGO3D outputs as ``ground truth'', and use the superscript $^{\mathrm{truth}}$ for them. Superscript $^{\mathrm{pred}}$ is for neural network predictions.
The hyperparameters used in our networks and training processes are listed in Table \ref{table:ml hyperparameters}.
\begin{deluxetable}{cc}

\tablecaption{Machine learning hyperparameters\label{table:ml hyperparameters}}

\tablehead{\colhead{Hyper-parameter} & \colhead{Default value}} 

\startdata
Training data size & 448 \\
Validation data size & 64 \\
Testing data size & 256 \\
Batch size & 32 \\
Num. steps for each batch & 3 \\
Initialization & Glorot normal \tablenotemark{a}\\
Learning rate & 0.0005 \\
Learning rate decay rate & 0.9 \\
Learning rate transition steps & 2000 \\
Optimizer & Adam \tablenotemark{b}\\
Total number of iterations & $10^5$ \\
Branch network layer size & 100, 100, 100, 100, 50 \\
Trunk network layer size & 256, 256, 256, 256, 256, 50 \\
Z network layer size & 100, 1 \\
\enddata
\tablenotetext{a}{\cite{glorot2010understanding}}
\tablenotetext{b}{\cite{kingma2014adam}}
\end{deluxetable} 

\section{Tests}\label{sec: tests}
We present a series of tests. In \S \ref{sec: 2d maps}, we compare the 2D maps of predicted surface density and velocity with the ground truth generated by \fargo{}.
Then we compare the predicted gap profile to the true profile in \S \ref{sec: 1d gap profile}. We further examine the behavior of our prediction on groups of disks by reproducing several empirical relationships from previous works in \S \ref{sec: test relation}.
    
    \subsection{2D maps of surface density and velocities}\label{sec: 2d maps}
\begin{figure*}
    \centering
    \includegraphics[trim=0 0 0 0, clip,width=0.95\textwidth,angle=0]{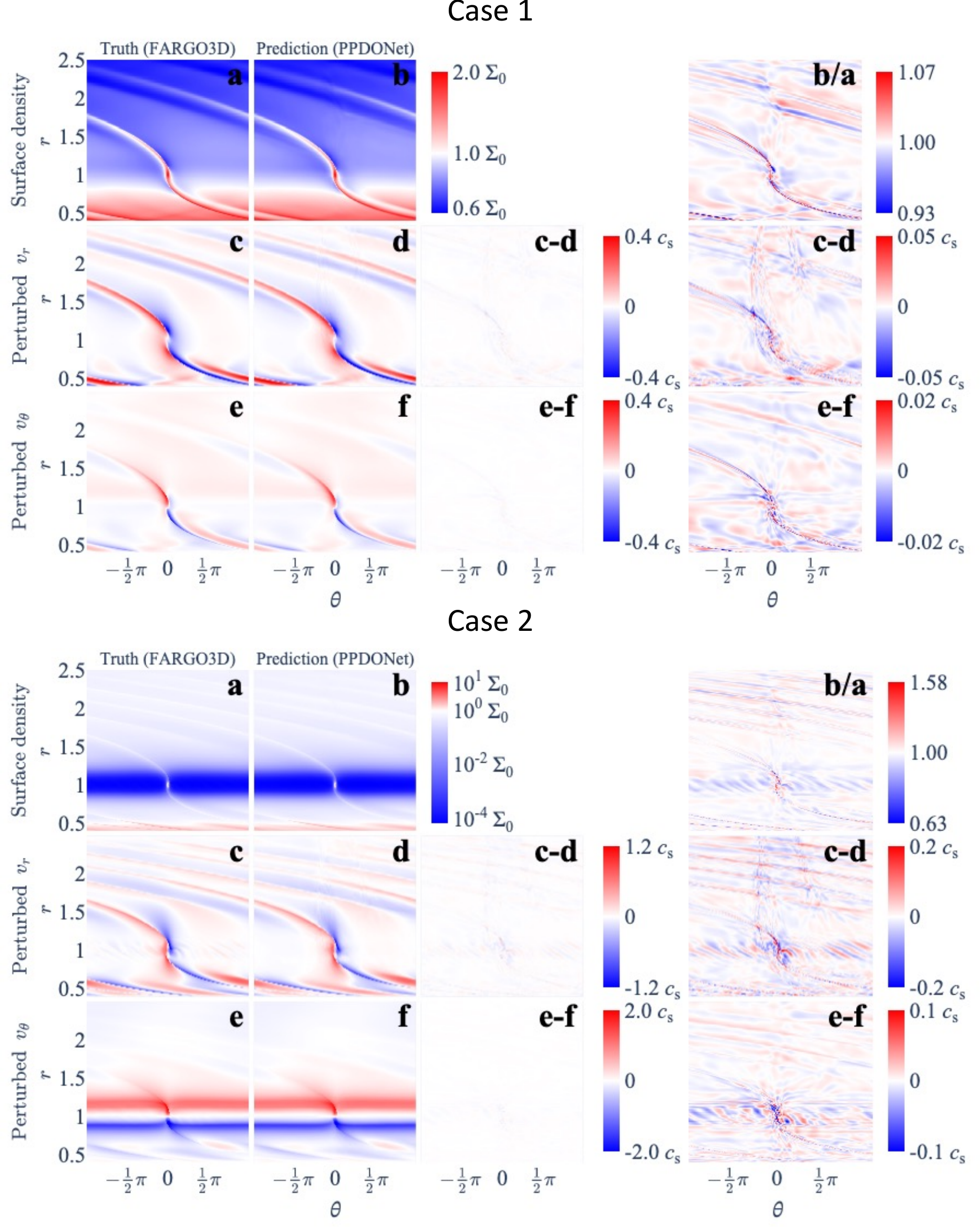}
    \caption{
    Two-dimensional comparisons of predicted surface density and velocity distribution. Case 1: $\left(\alpha,h_\mathrm{0},\pmass{}\right)=\left(0.013,0.092,6.0\times10^{-4}\right)$. Case 2: $\left(\alpha,h_\mathrm{0},\pmass{}\right)=\left(5.2\times10^{-4},0.053,1.6\times10^{-3}\right)$. 
    Left (a, c, and e): Ground truth generated from \fargo{} simulations. Middle (b, d, and f): Neural network predictions. For the two left columns, we subtract the initial value from the two velocities to highlight the perturbations. Two right columns: Differences or ratios. Surface density differences are measured by the ratio $\Sigma^{\mathrm{pred}}/\Sigma^{\mathrm{truth}}$, so that the gap region is highlighted. Velocity differences are shown in absolute errors.
    }
    
    \label{fig: test 2d maps}
\end{figure*}

Two representative cases are presented in Figure \ref{fig: test 2d maps}, highlighting density waves (case 1 with $\alpha=0.013$, $\ar{}=0.092$, and $\pmass{}=6.0\times10^{-4}$) and gaps (case 2 with $\alpha=5.2\times10^{-4}$, $\ar{}=0.053$, and $\pmass{}=1.6\times10^{-3}$). The left column shows the ground truth from \fargo{}, the middle column shows the PPDONet predictions, and the two right columns show the differences or ratios between the two. To ease the comparison in the perturbed radial and azimuthal velocity, we subtract their initial values from the corresponding panels. 
For surface density, we calculate the ratio between the prediction and ground truth to highlight the gap region. For velocities, which can take both positive and negative values, we show absolute errors.

A visual inspection of the model-prediction comparisons (panels $a,c,e$ {\it vs} $b,d,f$) shows that the differences in the location and contrast of both density waves and gaps are negligible. This is achieved despite these perturbations' sharp morphology relative to the background.

\subsection{1D surface density profile}\label{sec: 1d gap profile}
\begin{figure}[htb!]
    \centering
    \includegraphics[trim=0 0 0 0, clip,width=0.5\textwidth,angle=0]{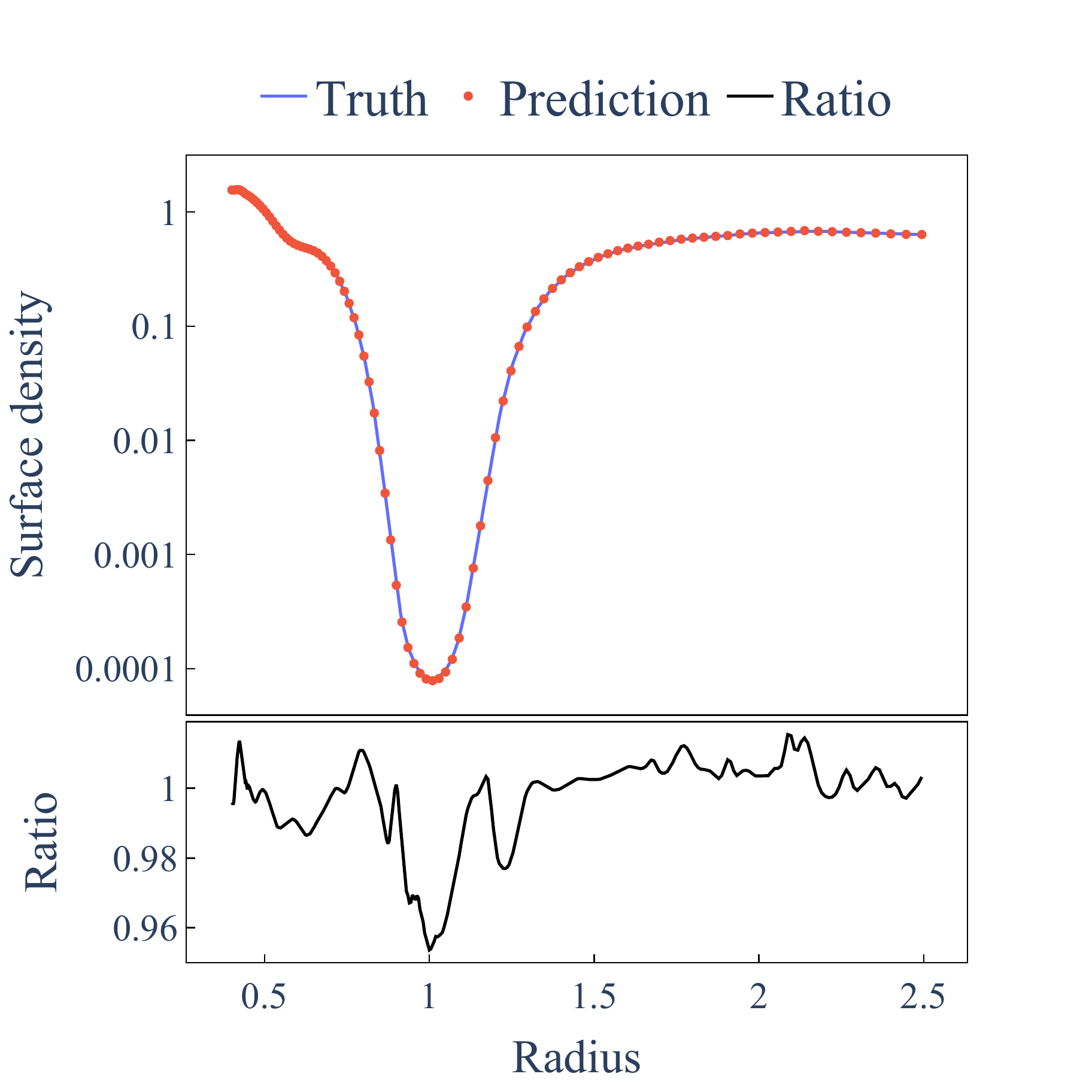}
    \caption{Gap profile comparison for one example case with $\alpha=5.2\times10^{-4}$, $\ar{}=0.053$, and $\pmass{}=1.6\times10^{-3}$. Blue: Ground truth from FARGO3D. Red: PPDONet prediction. Bottom panel: The ratio between the two. 
    }
    \label{fig: gap profile comparison}
\end{figure}
Gaps are one of the most important disk structures whose profiles are closely related to and can be used to constrain the properties of gap-opening planets \citep{kanagawa2016mass}. We carry out quantitative comparisons between PPDONet predictions and the ground truth by analyzing the azimuthally averaged 1D surface density radial profile of gaps. To measure 1D profiles, we mask regions contaminated by the planet and azimuthally average the surface density. One representative example ($\alpha=5.2\times10^{-4}$, $\ar{}=0.053$, and $\pmass{}=1.6\times10^{-3}$) is shown in Figure \ref{fig: gap profile comparison}.
The gap profiles from \fargo{} and PPDONet prediction overlap with no noticeable difference, and the ratio between the two deviates from unity by $\sim$1\%, indicating an excellent agreement.

\subsection{Empirical dependence of the morphology of disk structures on disk and planet properties}\label{sec: test relation}
\begin{figure}
    \centering
    \includegraphics[trim=0 0 0 0, clip,width=0.5\textwidth,angle=0]{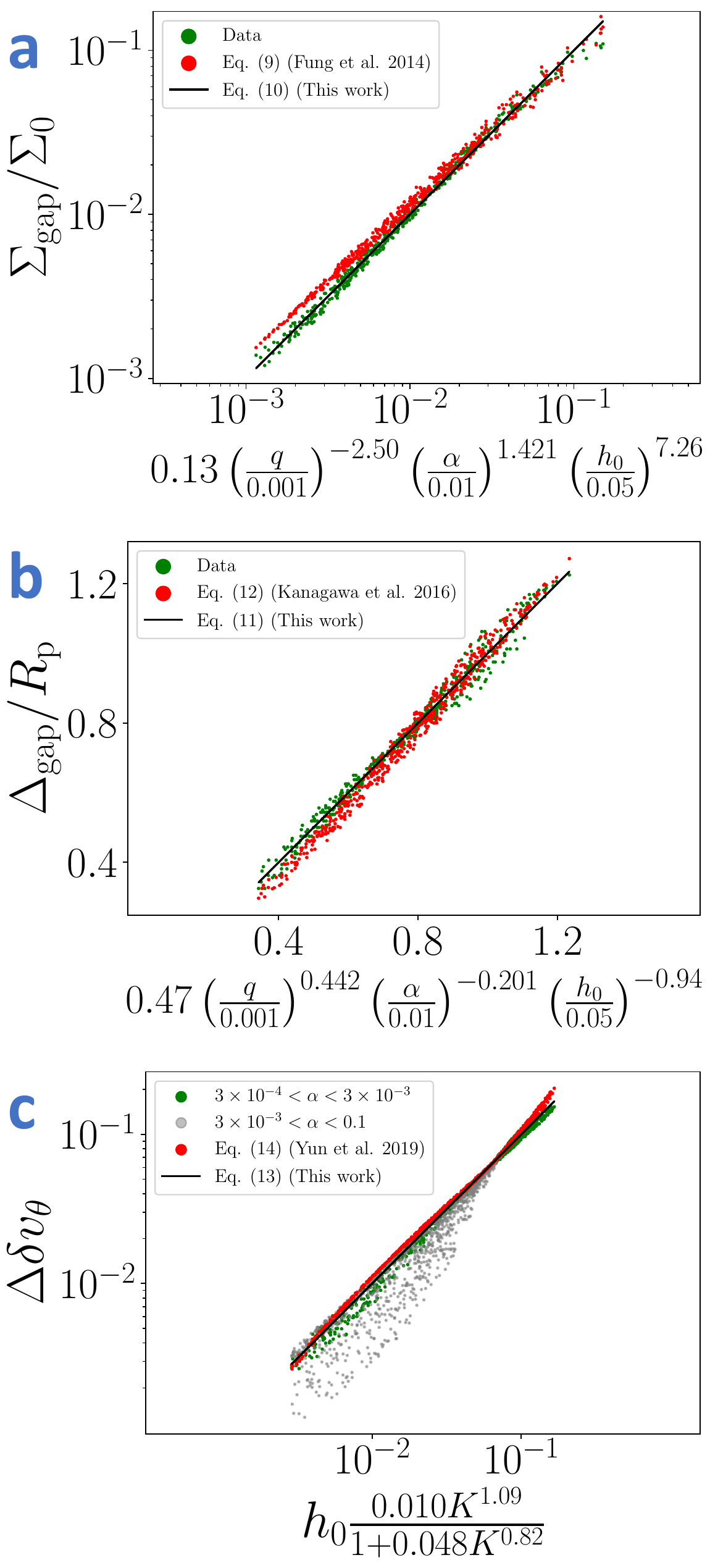}
    \caption{
    Empirical relationships for disk morphology. (a) presents the gap depth $\Sigma_\mathrm{gap}$ as a function of $\alpha$, $\ar{}$, and $\pmass{}$; (b) shows the gap width $\Delta_\mathrm{gap}$ against $\alpha$, $\ar{}$, and $\pmass{}$; and (c) depicts the amplitudes of perturbed rotational velocity $\deltav$ against the dimensionless parameter $K\equiv\pmass{}^2 \ar{}^{-5}\alpha^{-1}$.}
    \label{fig: relation}
\end{figure}

A few quantitative empirical relationships have been synthesized to connect the properties of disk features produced by planets to the properties of planets and the disk. We examine three such relationships using PPDONet-simulated disks, generated within ten minutes on a laptop.

\paragraph{Gap Depth ($\Sigma_\mathrm{gap}$).} The average surface density inside a gap depends on $\alpha$, $\ar{}$, and $\pmass{}$. \cite{fung2014empty} simulated $21$ disk-planet systems and fitted an empirical relationship when the planet masses were less than five times that of Jupiter:
\begin{equation}
    \Sigma_\mathrm{gap}/\Sigma_\mathrm{0}=0.14\left(\frac{\pmass{}}{0.001}\right)^{-2.16}\left(\frac{\alpha}{0.01}\right)^{1.41}\left(\frac{\ar{}}{0.05}\right)^{6.61}\label{eq: fung}.
\end{equation}
We measure gap depth in our sample of $500$ disk models with the same method as \cite{fung2014empty}, namely averaging the surface density inside the annulus $|r-r_\mathrm{p}|<2 {\rm max}(R_\mathrm{H},h)$, excluding the region $|\phi-\phi_\mathrm{p}|<2 {\rm max}(R_\mathrm{H},h)/r_\mathrm{p}$.
We fit the data with a power-law similar to \cite{fung2014empty} and measure the standard errors of the estimate, namely standard deviation of the error terms, using IBM SPSS Statistics. We get an adjusted $R^2$ equaling $0.995$, indicating an excellent fit
\begin{eqnarray}
    \Sigma_\mathrm{gap}/\Sigma_\mathrm{0}
    &=&0.13\pm{0.02}\left(\frac{\pmass{}}{0.001}\right)^{-2.50\pm{0.01}}\nonumber\\
    &\times&\left(\frac{\alpha}{0.01}\right)^{1.421\pm{0.007}}\left(\frac{\ar{}}{0.05}\right)^{7.26\pm{0.03}}\label{eq: gap depth}.
\end{eqnarray}
Figure \ref{fig: relation}a shows the data, the prediction of Equation \eqref{eq: fung}, and our fit. Note that due to the construction of the horizontal axis, the prediction of Equation \eqref{eq: fung} is shown using dots instead of a curve. While the power-law indexes of our fit differ slightly from those of \cite{fung2014empty}, the two exhibit the same trend: gaps are deeper in disks with lower viscosity and aspect ratio opened by more massive planets. We note that with hundreds of disk models, we are not only able to find a fitting function, but also to constrain the {\it uncertainties} in the fitted parameters, which are hard to achieve with tens of disk models produced using numerical solvers.

 \paragraph{Gap Width ($\Delta_{\rm gap}$)}. Gap width also relates to disk and planet properties and is an observable frequently used to constrain planet masses in observations \citep{kanagawa2016mass,zhang2018disk}. We measure the gap width in our sample and examine the empirical relationship in \cite{kanagawa2016mass}. We adopt a similar method in measuring $\Delta_{\rm gap}$ as \cite{kanagawa2016mass}, i.e., the radial separation between the inner and outer gap edges, with the edges identified at $0.5\ \Sigma_0$. We fit the data by a power law:
\begin{eqnarray}
    \Delta_\mathrm{gap}/R_\mathrm{p}
    &=&0.47\pm{0.04}\left(\frac{\pmass{}}{0.001}\right)^{0.442\pm{0.003}}\nonumber\\
    &\times&\left(\frac{\alpha}{0.01}\right)^{-0.201\pm{0.002}}\left(\frac{\ar{}}{0.05}\right)^{-0.94\pm{0.02}}.
\end{eqnarray}
In comparison, Equation (4) in \cite{kanagawa2016mass} gives
\begin{equation}
    \Delta_\mathrm{gap}/R_\mathrm{p}=0.39\left(\frac{\pmass{}}{0.001}\right)^{0.5}\left(\frac{\alpha}{0.01}\right)^{-0.25}\left(\frac{\ar{}}{0.05}\right)^{-0.75}.
\end{equation}
The two are shown in Figure \ref{fig: relation}b, a reproduction of Figure 3 in \cite{kanagawa2016mass}, with $19\times$ more data points ($500$ in our case {\it vs} $26$ in \cite{kanagawa2016mass}).
Again, the more than one order of magnitude bigger sample size enables us to obtain uncertainties in the fitted parameters.

\paragraph{Azimuthal Velocity Perturbation ($\deltav$).} 
We also examine the azimuthally averaged normalized perturbed rotational velocity $v_\theta(r,\theta)$: $\deltavr\equiv\left\langle\left(v_\theta(r,\theta)-v_{\theta, 0}(r,\theta)\right) / v_{\theta, 0}(r,\theta)\right\rangle$, where $v_{\theta, 0}(r,\theta)$ is the initial value. 
We then define the difference between the maximum and minimum $\deltavr$ within a radius range of $r\in(0.5 r_\mathrm{p}, 1.5 r_\mathrm{p})$ as
$\deltav\equiv\max\left[\deltavr\right]-\min\left[\deltavr\right]$
\footnote{The $\deltav$ in our work is defined as $\delta_{V}$ in \cite{yun2019properties}.} \cite[see][Figure~1d]{yun2019properties}.
The measurements are conducted on two datasets each containing $500$ disks in different parameter spaces. 
First, we generate disks from PPDONet with the same range of $\alpha{}$ as \cite{yun2019properties}: $3\times10^{-4}<\alpha<3\times10^{-3}$. We measure the perturbed velocity amplitudes in our sample (green dots) following \cite{yun2019properties} and obtain the best fit (black line):
\begin{equation}
    \deltav=\ar{}\frac{(0.010\pm{0.001})K^{1.09\pm{0.03}}}{1+(0.048\pm{0.003})K^{0.82\pm{0.02}}},
\end{equation}
where $K\equiv\pmass{}^2 \ar{}^{-5}\alpha^{-1}$. This is similar to the one obtained by (red dots) \cite{yun2019properties}
\begin{equation}
    \deltav=\ar{}\frac{0.007K^{1.38}}{1+0.06K^{1.03}}.
\end{equation}
The data and the two fits are shown in Figure \ref{fig: relation}c.
Next, we generate disks from PPDONet with larger $\alpha$, $3\times10^{-3}<\alpha<0.1$, and measure $\deltav$ (grey dots). We find neither relationship is suitable for accurately characterizing disks with high viscosity. 
This is as expected, because when $\alpha\gtrsim0.01$ viscous damping of density waves strongly affects the evolution of waves and their profiles \citep{miranda2020gaps}.

\section{SUMMARY}\label{sec: summary}
We develop a machine learning tool, PPDONet, based on Deep Operator Networks \citep[DeepONets;][]{lu2021learning} to predict the steady-state solutions of disk-planet interactions in parameterized protoplanetary disks. 
We train PPDONet using $448$ disks generated by \fargo{} simulations. The trained operator is able to map three scalar parameters -- the Shakura \& Sunyaev viscosity $\alpha$, the disk aspect ratio $\ar{}$, and the planet-star mass ratio $\pmass{}$ -- to steady-state solutions of disk surface density, radial velocity, and azimuthal velocity. 
Currently, PPDONet is the first and only public tool that is able to accomplish this forward problem.
It can predict the structures of $500$ disk-planet systems in a few minutes on a laptop, many orders of magnitude faster than conventional numerical simulations.

To evaluate the performance of PPDONet, we present a comprehensive set of tests. We compare 2D maps (\S \ref{sec: 2d maps}) and 1D surface density radial profiles (\S \ref{sec: 1d gap profile}) generated from PPDONet with those produced by FARGO3D simulations. The two are consistent with each other with little noticeable difference. We use PPDONet to generate multiple samples of disk-planet interaction, each containing on the order of $1,000$ disks, and revisit several empirical relationships previously reported (\S \ref{sec: test relation}), including how gap depth, gap width, and azimuthal velocity perturbation depend on $\alpha$, $\ar{}$, and $\pmass{}$. Overall, we recover previously found correlations. In addition, thanks to the one to two orders of magnitude larger samples that PPDONet is able to quickly produce compared with those previously produced using conventional numerical solvers, we are able to constrain the uncertainties in the fitting parameters, a nearly impossible task in the past due to small sample sizes.

Our tool is positioned to replace conventional numerical simulations in certain applications. It is publicly available at \url{https://github.com/smao-astro/PPDONet}.

\section*{acknowledgments}
We are grateful to an anonymous referee for constructive suggestions that improved our paper. 
We thank Xuening Bai, Pablo Ben{\'\i}tez-Llambay, Shengze Cai, Miles Cranmer, Bin Dong, Xiaotian Gao, Jiequn Han, Pinaghui Huang, Xiaowei Jin, Hui Li, Tie-Yan Liu, Chris Ormel, Wenlei Shi, Karun Thanjavur, Yiwei Wang, Yinhao Wu, Zhenghao Xu, Minhao Zhang, and Wei Zhu for help and useful discussions in the project.

S.M. and R.D. are supported by the Natural Sciences and Engineering Research Council of Canada (NSERC) and the Alfred P. Sloan Foundation. S.M. and R.D. acknowledge the support of the Government of Canada's New Frontiers in Research Fund (NFRF), [NFRFE-2022-00159].
This research was enabled in part by support provided by the Digital Research Alliance of Canada \url{alliance.can.ca}.

%

\vspace{5mm}


\software{FARGO3D \citep{masset2000fargo,benitez2016fargo3d},
    NumPy \citep{harris2020array},
    xarray \citep{hoyer2017xarray},
    JAX \citep{jax2018github,deepmind2020jax},
    Haiku \citep{haiku2020github}
    }



\pagebreak
\appendix
\section{Training Dataset Size and Test Error}
\begin{figure*}[h]
    \centering
    \includegraphics[trim=0 0 0 0, clip,width=0.9\textwidth,angle=0]{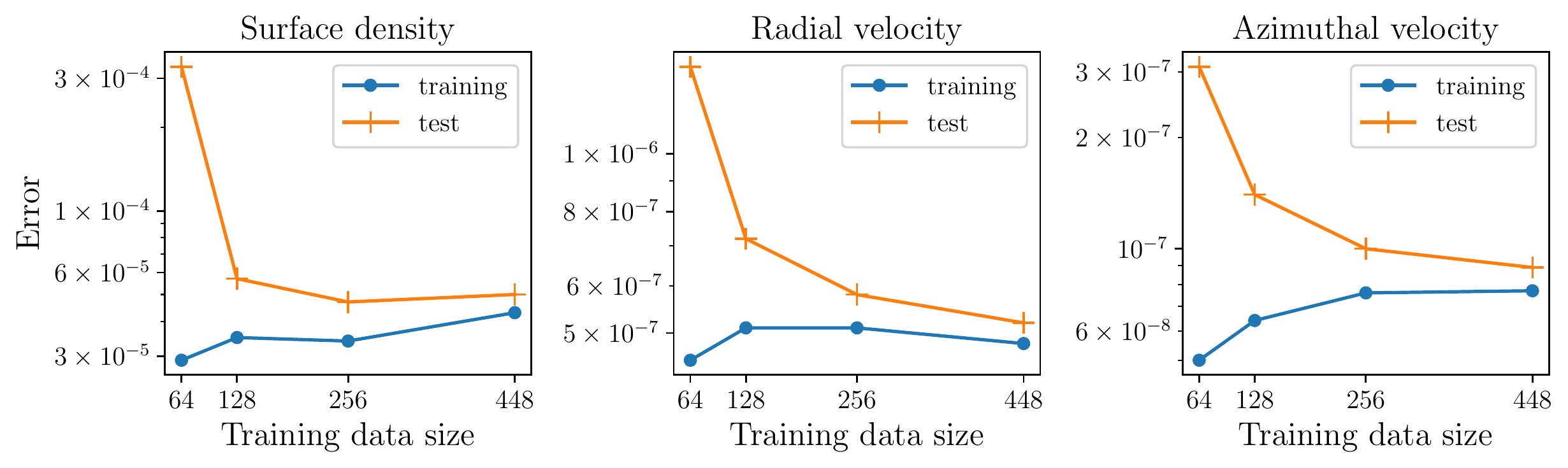}
    \caption{Comparison of test errors for neural networks trained using training dataset with different sizes.
    The three columns represent networks trained for surface density (left), radial velocity (middle), and azimuthal velocity (right), respectively.
    We calculate the losses using the method described in \S \ref{sec: network training}.
    }
    
    \label{fig: test dataset sizes 1}
\end{figure*}
\begin{figure*}[h]
    \centering
    \includegraphics[trim=0 0 0 0, clip,width=0.9\textwidth,angle=0]{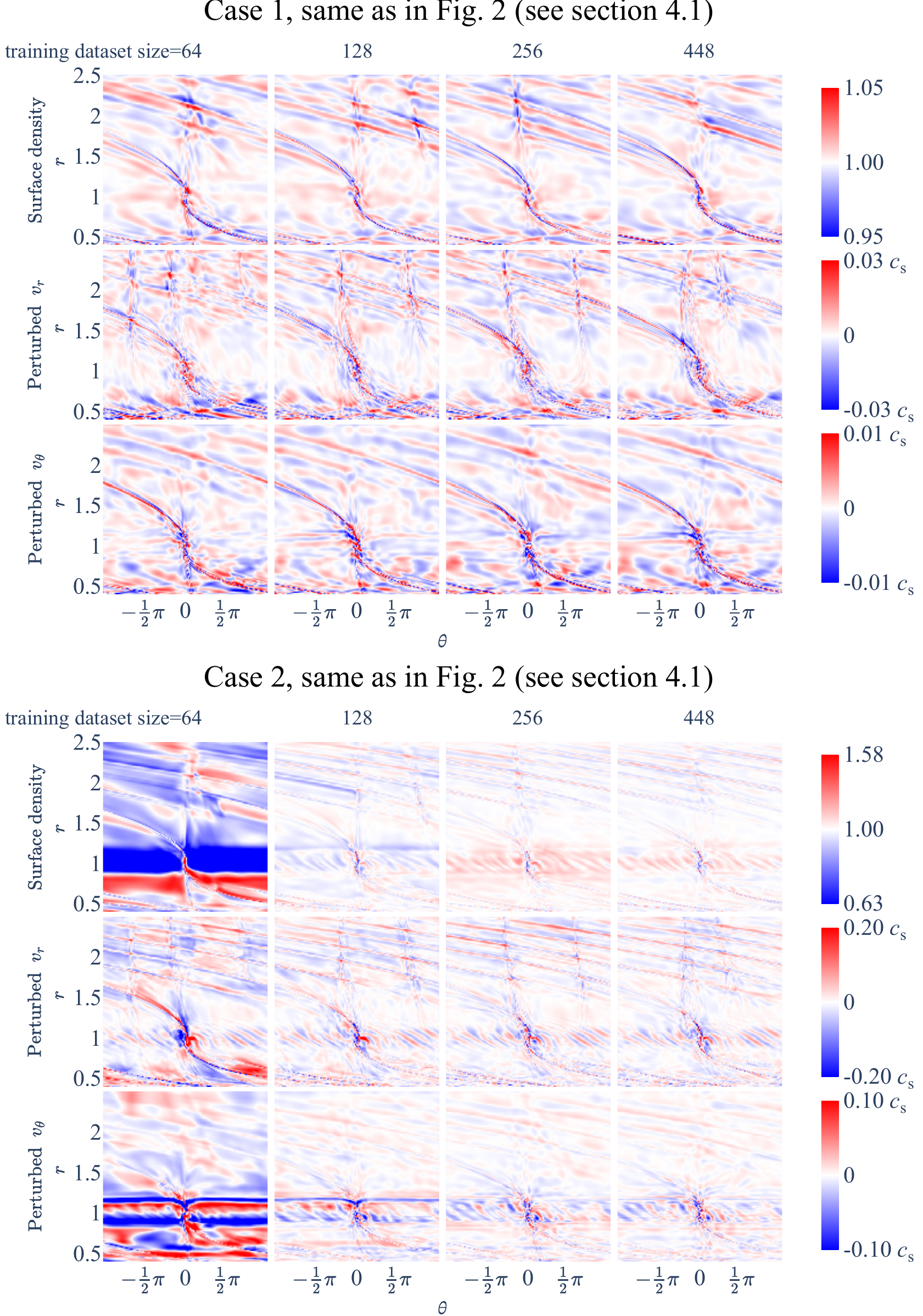}
    \caption{Comparison of prediction errors for neural networks trained with different dataset sizes.
    The four columns represent networks trained on dataset sizes of 64 (leftmost column), 128, 256, and 488 (rightmost column) samples, respectively.
    The representative cases and difference calculation methodology correspond to those utilized in Figure \ref{fig: test 2d maps}.
    }
    
    \label{fig: test dataset sizes 2}
\end{figure*}
The predictive accuracy of the network is contingent on the sizes of the training datasets \citep[Figure~4]{lu2021learning}, and can be problem dependent.
To investigate the data size dependency in our problem, we train twelve networks, each targeting surface density, radial velocity, and azimuthal velocity, using training datasets of sizes 64, 128, 256, and 448, respectively. The networks are then tested on a dataset comprising 256 unseen \fargo{} simulations (see \S~\ref{sec: hydro simulations}).

As the number of \fargo{} simulations in the training data increases, the test errors initially decrease before eventually plateauing due to factors such as limited network capacity (Figure~\ref{fig: test dataset sizes 1}). To provide a clearer visualization of the comparison, we plot the two-dimensional difference between the \fargo{} simulations and the network predictions in Figure~\ref{fig: test dataset sizes 2}. The representative cases and difference calculation methodology employed in this figure correspond to those used in Figure \ref{fig: test 2d maps}.




\end{document}